\documentclass[%
superscriptaddress,
nofootinbib,
 aps,
 twocolumn,
 prl
]{revtex4-1}

\usepackage{amsmath,amssymb}
\usepackage{wasysym}
\usepackage{graphicx}
\usepackage{slashed}
\usepackage{color}
\usepackage[symbol]{footmisc}
\usepackage{hyperref}

\usepackage{ulem}
\newcommand{\beq}{\begin{equation}}
\newcommand{\eeq}{\end{equation}}
\newcommand{\bea}{\begin{eqnarray}}
\newcommand{\eea}{\end{eqnarray}}

\allowdisplaybreaks
\begin{document}

\author{Zuo-tang Liang}
\email{liang@sdu.edu.cn}
\affiliation{Institute of Frontier and Interdisciplinary Science,
Key Laboratory of Particle Physics and Particle Irradiation (MOE),
Shandong University, (QingDao), Shandong 266237, China}

\author{Qian Yang}
\email{yangqian2020@sdu.edu.cn}
\affiliation{Institute of Frontier and Interdisciplinary Science,
Key Laboratory of Particle Physics and Particle Irradiation (MOE),
Shandong University, (QingDao), Shandong 266237, China}

\author{Hong Zhang}
\email{hong.zhang@sdu.edu.cn}
\affiliation{Institute of Frontier and Interdisciplinary Science,
Key Laboratory of Particle Physics and Particle Irradiation (MOE),
Shandong University, (QingDao), Shandong 266237, China}

\author{Jiaxing Zhao}
\email{jzhao@itp.uni-frankfurt.de}
\affiliation{Helmholtz Research Academy Hesse for FAIR (HFHF), GSI Helmholtz Center for Heavy Ion Physics, Campus Frankfurt, 60438 Frankfurt, Germany}
\affiliation{Institut fu\:r Theoretische Physik, Johann Wolfgang Goethe-Universität,Max-von-Laue-Straße 1, D-60438 Frankfurt am Main, Germany}
\affiliation{Department of Physics, Tsinghua University, Beijing 100084, China}

\author{Pengfei Zhuang}
\email{zhuangpf@mail.tsinghua.edu.cn}
\affiliation{Department of Physics, Tsinghua University, Beijing 100084, China}

\date{\today}

\title{Searching for True Muonium in Relativistic Heavy Ion Collisions}

\begin{abstract}
We investigate the production of the as-yet-undetected true muonium within the quark-gluon plasma formed in relativistic heavy-ion collisions, employing a relativistic Boltzmann transport framework coupled to viscous hydrodynamic simulations.
The obtained effective cross sections for central collisions are 1.23~$\mu b$ in AuAu collisions with $\sqrt{s_{\rm NN}}=200$~GeV and 14.2~$\mu b$  in PbPb collisions with $\sqrt{s_{\rm NN}}=5.02$~TeV, resulting in a yield of $\mathcal{O}(10^4)$ and $\mathcal{O}(10^5)$ true muonium per billion $AA$ collisions at RHIC and the LHC, respectively. 
This establishes heavy-ion collisions as a promising process for detecting true muonium.
%\newpage 
\end{abstract}

\maketitle

%%%%%%%%%%%%%%%%%%%%%%%%%%%%%%%%%%%%%%
{\it Introduction.} 
True muonium $(\mu^+\mu^-)$ is the hydrogen-like bound state of a muon and anti-muon pair, which provides a unique laboratory for testing fundamental quantum electrodynamics (QED) and probing lepton universality \cite{Bilenky:1969pw, Bilenky:1969zd, Malenfant:1987tf, Jentschura:1997tv, Jentschura:1998vkm,Ji:2016fat}. As the heaviest pure leptonic atom, its precise spectroscopy could reveal deviations from Standard Model predictions and provide a clean window into new physics \cite{Tucker-Smith:2010wdq,Lamm:2016jim}. Realizing this potential requires first overcoming the experimental challenge of producing and detecting this ephemeral state. The bound state's relatively small mass, short lifetime, and low production cross sections require either high-luminosity sources or large boost factors enabling secondary vertex detection. Current attempts mainly focus on four complementary approaches: direct production in high-energy colliders \cite{Crane:1971bx,Moffat:1975uw,Holvik:1986ty, Arteaga-Romero:2000mwd, Brodsky:2009gx, Fox:2021mdn, Gargiulo:2023tci, Gargiulo:2025pmu}, fixed-target experiments near threshold \cite{Arteaga-Romero:2000mwd,Banburski:2012tk,  Gargiulo:2024zyc}, rare decays of heavy hadrons \cite{Nemenov:1972ph, Vysotsky:1979nv, Malenfant:1987tm, Kozlov:1987ey, Ji:2017lyh, Ji:2017lyh, Ji:2018dwx,  CidVidal:2019qub} as well as true muonium formation via two- and three-photon fusion processes in ultra-peripheral heavy ion collisions \cite{Ginzburg:1998df, Azevedo:2019hqp, Yu:2013uka, Azevedo:2019hqp, Dai:2024imb}. With all these efforts, this bound state is still unobserved.

Heavy-ion collisions (HICs), characterized by their multi-body QCD interactions and large participant numbers, have established a unique discovery platform for exotic quantum states. This paradigm is exemplified by the historic detection of anti-nuclei up to anti-helium-4 \cite{STAR:antiH4L, STAR:antiH4}. The quark-gluon plasma (QGP) fireball created in relativistic HICs exhibits extraordinary properties - temperatures of few hundred MeV,  lowest known viscosity-to-entropy ratio\cite{QGP:WhitePaper}, and the most vortical fluid produced in the laboratory\cite{STAR:Lambda}. Particularly, the medium temperature is higher than the true muonium mass, suggesting the bound state could be produced thermally. Experimentally, the QGP thermal emissivity has been investigated with the characteristic photon\cite{Paul:DirectPhoton} and dilepton\cite{NA60:dimuon,STAR:dilepton1} spectra. The dielectron spectroscopy at the CERN's NA60+ campaign \cite{NA60:dimuon} and recently at RHIC's STAR experiment \cite{STAR:dilepton1} has also been employed as a powerful tool to measure the QGP temperature profiles. 

In this work, the thermal production of $n^3S_1$-state $(\mu^+ \mu^-)$ in QGP is investigated. Below we first analyze the characteristic time scales of true muonium and QGP produced at RHIC and the LHC, arguing that the QGP only serves as a source of the $\mu^+$ and $\mu^-$, while not interfering with the formation or decay of the bound state. Then we calculate the scattering amplitudes for two dominant processes $q\bar{q}\to(\mu^+ \mu^-)g$ and $q g\to(\mu^+ \mu^-)q$. Finally, these amplitudes are integrated into a relativistic Boltzmann transport framework, self-consistently coupled to viscous hydrodynamic simulations of QGP evolution. 
 
Our calculation differs from that in Ref.~\cite{Chen:2012ci} in two aspects. First, we have used the thermal quark masses with three flavors and temperature-dependent strong coupling. Secondly, the viscous hydrodynamic simulation offers a more accurate description of the medium than the ideal hydrodynamic equations. These two improvements turn out to be essential. Our results predict $n^3S_1$ true muonium yields of approximately $\mathcal{O}(10^4)$ and $\mathcal{O}(10^5)$ per billion central collisions at RHIC and the LHC, respectively. These values are more than $10^3$ times larger than the counterparts obtained in Ref.~\cite{Chen:2012ci}, allowing a discovery of $(\mu^+\mu^-)$ at RHIC and the LHC.

{\it Characteristic Scales.} 
The production of true muonium in QGP is governed by their distinct temporal scales. In this work, we focus on the $n^3S_1$ state.  The true muonium $(\mu^+ \mu^-)$ in a vacuum can be considered as a pure QED state with a lifetime of $\mathcal{O}(1)$~ps, six decades smaller than the muon lifetime  \cite{Jentschura:1998vkm}. It mainly decays to $e^+e^-$ with a branching ratio $98\%$. The binding energy is $E_b\approx -1.4/n^2$~keV and the mass is $m_n\approx 2 m_\mu=211$~MeV, where $m_\mu$ is the muon mass. The mean radius of the bound state can be estimated with $r_b\sim(m_\mu E_b)^{1/2}=512\,n$~fm.  The bound state formation time is then $\tau_b\sim r_b/c=1.72\times 10^{-21}$~sec. Comparably, the maximum temperature of the QGP formed at RHIC or LHC can reach hundreds of MeV, with a typical lifetime of $\tau_\text{QGP} \sim\mathcal{O}(10^{-23})$~sec and a typical size of about 10~fm. In the presence of QGP, the interaction potential between the $\mu^+\mu^-$ pair is Coulomb-like outside the fireball but Yukawa-like $e^{-r/r_D}/r$ in the QGP due to Debye screening, where $r_D\approx 2.11 (k_B T)^{-1}$ is the photon Debye length. The modified binding energy is larger than $-1.4$~keV, increasing with the QGP size $R_\text{QGP}$ and temperature $T$. For $T=500$~MeV and $R_\text{QGP}=15$~fm, the difference is only 1.5~eV, which could be safely ignored compared to the value of $E_b$. Moreover, the lifetime of true muonium is also unaffected by the presence of QGP, since $(\mu^+\mu^-)\to q\bar{q}$ process is forbidden with the thermal quark mass. Because $\tau_b\gg \tau_\text{QGP}$ and $r_b\gg r_\text{QGP}$, the QGP does not interfere with the formation of the bound state. 

{\it Production process.} 
The constituents in the strongly-interacting QGP are massive effective quasiparticles with broad spectral functions. Here we adopt the Dynamical Quasiparticle Model (DQPM), in which the partons' thermal masses and the widths of these quasiparticles as well as the strong coupling are extracted by comparing the dynamical quasiparticle entropy density $s$ with the lattice calculation \cite{Berrehrah:2013mua}. The dressed masses for the gluon $g$, light quarks $q=u,d$, and strange quark $s$ are,
%===
\begin{subequations}
\begin{align}
m_g^2(T)&=\frac{g^2(T)}{6}\left(N_c+\frac{1}{2}N_f\right)T^2,\\
%%%
m_q^2(T)&=\frac{N_c^2-1}{8N_c}g^2(T)T^2,\\
%%%
m_s(T)&=m_q(T)+0.045~\text{GeV},
\end{align}
\end{subequations}
%===
where $N_c=N_f=3$. The strong coupling $g(T)$ is fitted with a piecewise function,
%===
\begin{align}
g^2(T)=\left\{
\begin{array}{ll}
\frac{48\pi^2}{(11N_c-2N_f)\log\left[\lambda^2\left(\frac{T}{T_c}-0.56\right)^2\right]}, & \hspace{0cm} T\geq T^*, \\
g^2(T^*)\left(\frac{T^*}{T}\right)^{3.1}, & \hspace{0cm} T<T^*,
\end{array}\right.
\end{align}
%=== 
where $T_c = 158~\text{MeV}$, $T^*=1.19\,T_c$ and $\lambda=2.42$. The finite width and the chemical potentials are also obtained in Ref.~\cite{Berrehrah:2013mua}. Their effects are studied and shown to be not important in the same reference. Below we ignore the finite width and the chemical potentials for simplicity.

%-------------------------------------------------------------------
\begin{figure}
\begin{center}
\includegraphics[width=0.3\textwidth]{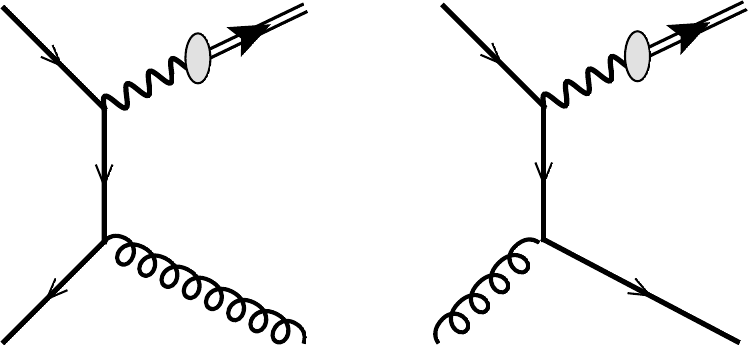}
\end{center}
\caption{Schematic Feynman diagrams of NLO processes.
}\label{fig:FD}
\end{figure} 
%-------------------------------------------------------------------

%-------------------------------------------------------------------
\begin{figure}
\begin{center}
\includegraphics[width=0.45\textwidth]{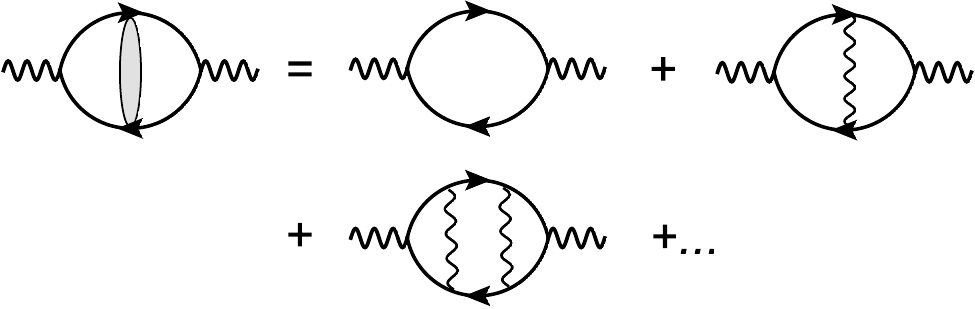}
\end{center}
\caption{Feynman diagrams for the bubble amplitude with Coulomb resummation.
}\label{fig:bubble}
\end{figure} 
%-------------------------------------------------------------------

With the dressed quark mass, the LO process $q\bar{q}\to (\mu^+\mu^-)$ is kinematically forbidden. There are two next-to-leading-order (NLO) processes: $q+\bar{q}\to (\mu^+\mu^-)+g$ and $q+g\to (\mu^+\mu^-)+q$, with example Feynmann diagrams showed in Fig.~\ref{fig:FD}. The coupling of the virtual photon to the $(\mu^+\mu^-)$ can be calculated by summing the photons exchanged between the nonrelativistic $\mu^+\mu^-$ pair in Fig.~\ref{fig:bubble} \cite{Kong:1998sx, Kong:1999sf}. The obtained photon-($\mu^+\mu^-$) vertex is
%===
\begin{align}
i\mathcal{Z}_n = i\frac{\alpha^2}{4\pi}\sqrt{\frac{m_\mu^3 m_n}{ n^3}},
\end{align}
%===
where $m_n$ is the mass of ($\mu^+\mu^-$) in the $n^3S_1$ state. $\alpha=1/137$ is the electromagnetic coupling constant. This approach is mathematically the same as mapping the $\mu^+\mu^-$ to the wavefunction at the center of the bound state widely used in literature \cite{Ginzburg:1998df,Chen:2012ci,Gargiulo:2024zyc}.

For $q(p_1)\bar{q}(p_2)\to (\mu^+\mu^-)(k_1) g(k_2)$ process, where in each bracket is the momentum of the corresponding particle, the squared amplitude is
%===
\begin{eqnarray}\label{eq:M1-sq}
\sum_{s,s',\lambda,i}|\mathcal{M}_1|^2 &=&
\frac{16\pi e_f^2 \alpha^4\alpha_s (N_c^2-1)m_\mu^3}{n^3 m_n^3}\nonumber\\
&\times&
\bigg[
-(m_n^2+2m_f^2)(m_g^2+2m_f^2)
\left(\frac{1}{t_f^2}+\frac{1}{u_f^2}\right)\nonumber\\
&+&\frac{u_f}{t_f}+\frac{t_f}{u_f}+(2(m_n^2+m_g^2)^2-8m_f^4) \frac{1}{t_f u_f}\nonumber\\
&-&2(m_n^2+m_g^2+2m_f^2)\left(\frac{1}{t_f}+\frac{1}{u_f}\right)
\bigg].
\end{eqnarray}
%===
Here we have summed the quark spin $s$, antiquark spin $s'$, gluon polarization $\lambda$, and color indices collectively labeled as $i$. Only the transverse gluon polarizations are considered in consistency with Ref.~\cite{Berrehrah:2013mua}. The label $f=u,d,s$ denotes the quark with charge $e_f$. The strong coupling constant is $\alpha_s=g^2(T)/(4\pi)$. The kinematic variables are $t_q=(p_1-k_1)^2-m_f^2$ and $u_q=(p_2-k_1)^2-m_f^2$. 

The calculation of the $q(p_1)g(p_2)\to (\mu^+\mu^-)(k_1) q(k_2)$ is very similar. By applying the crossing symmetry, the squared amplitude is Eq.~\eqref{eq:M1-sq} with $u_f$ replaced by $s_f=(p_1+p_2)^2-m_f^2$. 

%%%%%%%%%%%%%%%%
%%%%%%%%%%%%%%%%

{\it Production in heavy ion collisions.}
In the relativistic heavy ion collisions, the produced QGP is not a static medium but a rapidly expanding system. Because the true muonium never reaches thermal equilibrium with the QGP medium, the accurate description of the true muonium production requires a transport approach like the Boltzmann equation. Due to its small phase space distribution and relatively long lifetime, one can safely ignore the dissociation of the $(\mu^+\mu^-)$ in the Boltzmann equation, which gives, 
%===
\begin{align}
\label{eq:BEq}
\left({\partial \over \partial t}+{{\bf p}\over E}{\partial \over \partial {\bf r}} \right) f(k_1)=  \sum_{i=1,2}C_{\text{NLO},i}[f(k_1)],
\end{align} 
%===
where the left-hand side is the diffusion term, and the right-hand side is the collision term.
For the $q\bar q\to (\mu^+\mu^-)+g$ process, the collision term can be expressed as,
%====
\begin{eqnarray}\label{eq:C-nlo-1}
C_{\text{NLO},1}&=&\frac{1}{E_{k_1}}\sum_{f,n}\int\frac{d^3 p_1}{(2\pi)^3 2E_{p_1}}\frac{d^3 p_2}{(2\pi)^3 2E_{p_2}}\frac{d^3 k_2}{(2\pi)^3 2E_{k_2}} \nonumber\\
&\times &\sum_{s,s',\lambda,i}|\mathcal{M}_1|^2 f_q(p_1) f_{\bar{q}}(p_2)\left(1+f_g(k_2)\right)\nonumber\\
&\times &(2\pi)^4\delta^{(4)}(p_1+p_2-k_1-k_2),
\end{eqnarray}
%====
where $f_q$, $f_{\bar{q}}$ and $f_g$ are the thermal distributions of the dressed quark, antiquark and gluons, respectively. Since the mass difference of $(\mu^+\mu^-)$ with different $n$ is much smaller than the detector resolution, they are considered collectively with $n$ summed. The approximation $m_n=2m_\mu$ further allows a factorization of the $n$-dependence as $\zeta_3=\sum_n n^{-3}$. The summation over the quark flavor $f$ runs over $u,d$ and $s$. The collision term for the $q g\to (\mu^+\mu^-)+q$ process, $C_{\text{NLO},2}$, could be obtained by changing the distributions in Eq.~\eqref{eq:C-nlo-1} to $f_q(p_1)f_g(p_2)\left(1-f_q(k_2)\right)$ and summing $f$ over $u,d,s$ and their antiparticles.

 %-------------------------------------------------------------------
\begin{figure}
\begin{center}
\includegraphics[width=0.45\textwidth]{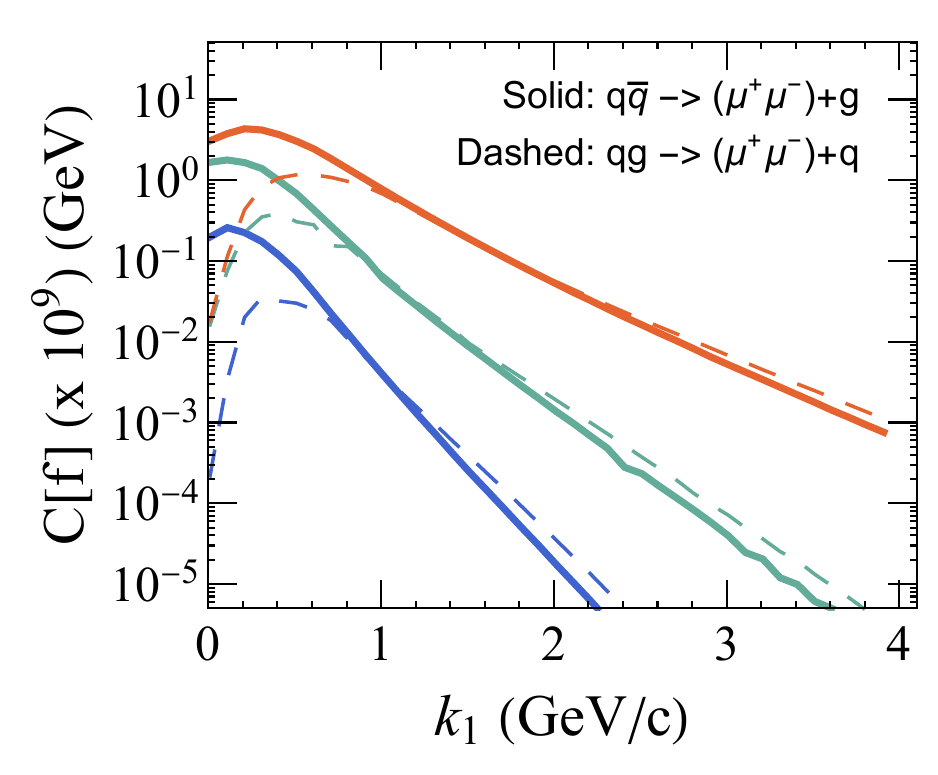}\\
\includegraphics[width=0.45\textwidth]{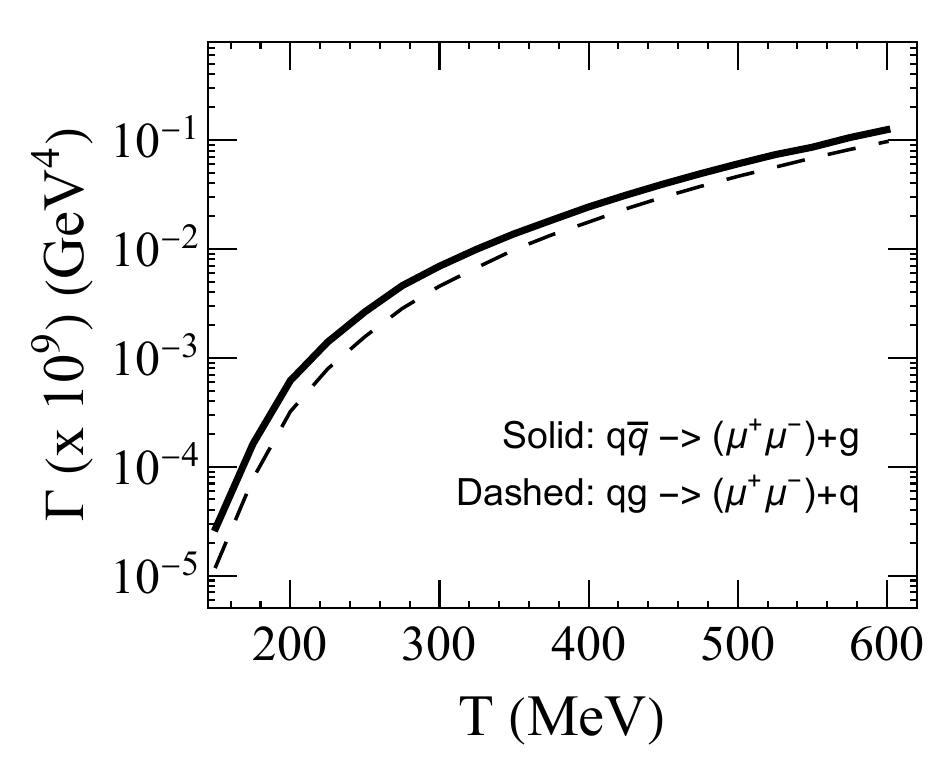}
\end{center}
\caption{The collision terms as functions of true muonium momentum $k_1$ with different temperature $T$ (top panel) and temperature-dependent production rates (bottom panel). The solid and dashed curves are the results of $q\bar q\to (\mu^+\mu^-)+g$ and $q g\to (\mu^+\mu^-)+q$ channel, respectively. Orange, green, and blue curves in the top panel are with different temperatures, $T=500,300,200~\rm MeV$.   
}\label{fig:collision}
\end{figure} 
 %-------------------------------------------------------------------
 
Fig.~\ref{fig:collision} shows the two NLO collision terms as functions of the true muonium momentum with different temperatures $T$. The $q\bar q\to (\mu^+\mu^-)+g$ channel dominates in the low momentum region but is comparable to the $q g\to (\mu^+\mu^-)+q$ channel in the high momentum region. The production rates defined as $\Gamma\equiv \int C_{\text{NLO}} d{\bf k}_1/(2\pi)^3$ are also plotted in Fig.~\ref{fig:collision} to study the temperature dependence. It is clear that the production rate quickly increases with the temperature.

The above Boltzmann equation, Eq.~\eqref{eq:BEq}, has an analytic solution~\cite{Chen:2012ci,Zhao:2020jqu},
%===
\begin{eqnarray}
f({\bf p}_T,y,{\bf x}_T,\eta,\tau)&=&\int_{\tau_0}^{\tau}d{\tau'} {C_{\rm NLO}[f({\bf p}_T,y,{\bf X}_T,H,\tau')] \over \cosh(\Delta(y-\eta))}\nonumber\\
&\times& \Theta(T({\bf X}_T, H, \tau')-T_c),
\label{eq.bolt.solu}
\end{eqnarray}
%=== 
with 
%===
\begin{eqnarray}
{\bf X}_T&=&{\bf x}_T-{\bf v}_T [\tau \cosh(y-\eta)-\tau' \cosh(\Delta)], \nonumber\\
H&=&y-\Delta, \nonumber\\
\Delta&\equiv &\text{arcsinh} \Big({\tau\over \tau'}\sinh(y-\eta) \Big).
\end{eqnarray} 
%===
The step function $\Theta$ indicates that the true muonium can only be generated in the deconfined phase with $T>T_c$. The variable ${\bf v}_T={\bf p}_T/E_T$ with $E_T=\sqrt{m^2+p_T^2}$ is the transverse energy. Then the yield at any time $\tau$ can be obtained by integrating on the hypersurface via,
%===
\begin{eqnarray}
N(\tau)&=&{1\over (2\pi)^3} \int d^2{\bf p}_T d^2{\bf x}_Tdy d\eta \tau E_T \cosh(y-\eta) \nonumber\\
&\times&f({\bf p}_T,y,{\bf x}_T,\eta,\tau).
\end{eqnarray} 
%===

 %-------------------------------------------------------------------
\begin{figure}%[!htb]
\begin{center}
\includegraphics[width=0.45\textwidth]{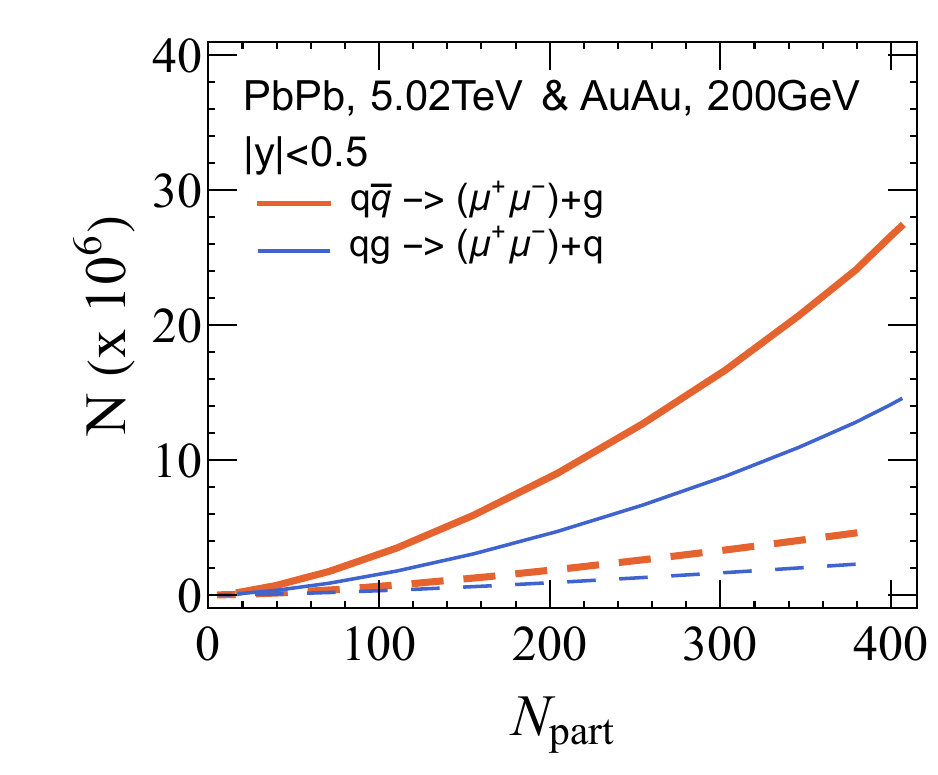}\\
\includegraphics[width=0.45\textwidth]{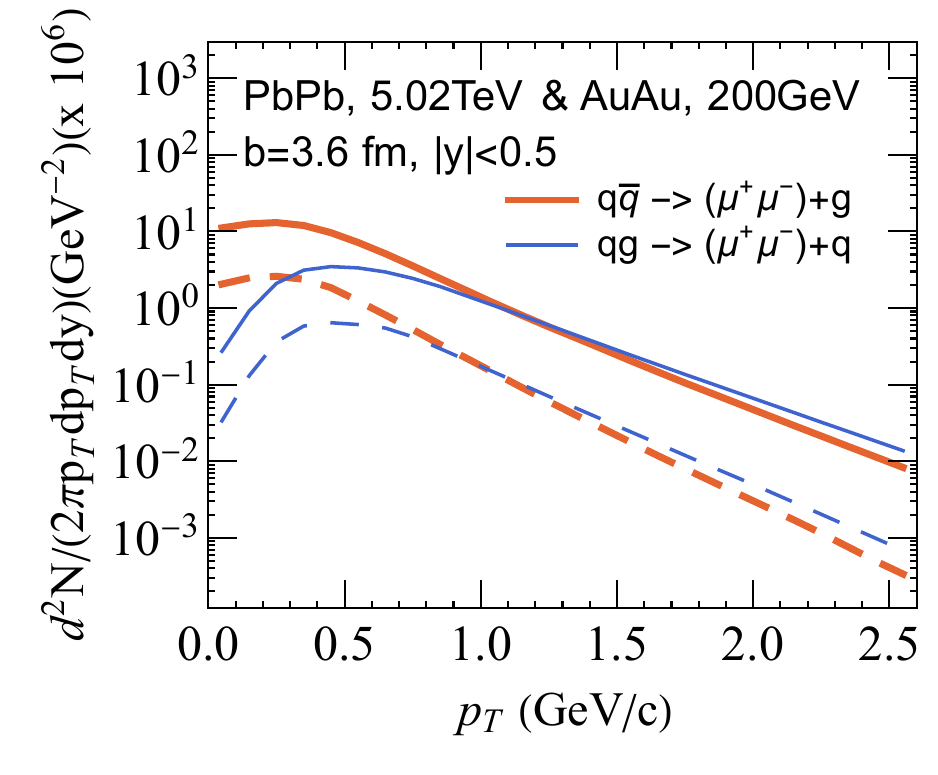}
\end{center}
\caption{The true muonium yield as a function of the number of participants $N_{\rm part}$ (top panel) and transverse momentum (bottom panel) in the central rapidity of Au+Au collisions at RHIC energy $\sqrt{s_{\rm NN}}$ = 200 GeV (dashed curves) and Pb+Pb collisions at LHC energy $\sqrt{s_{\rm NN}}$ = 5.02 TeV (solid curves). The orange lines show the results from the $q\bar q\to (\mu^+\mu^-)+g$ channel, while blue lines show the results from the $q g\to (\mu^+\mu^-)+q$ channels. }
\label{fig.yield-pt}
\end{figure}
%--------------------------end------------------------------------

 %-------------------------------------------------------------------
\begin{figure}%[!htb]
\begin{center}
\includegraphics[width=0.45\textwidth]{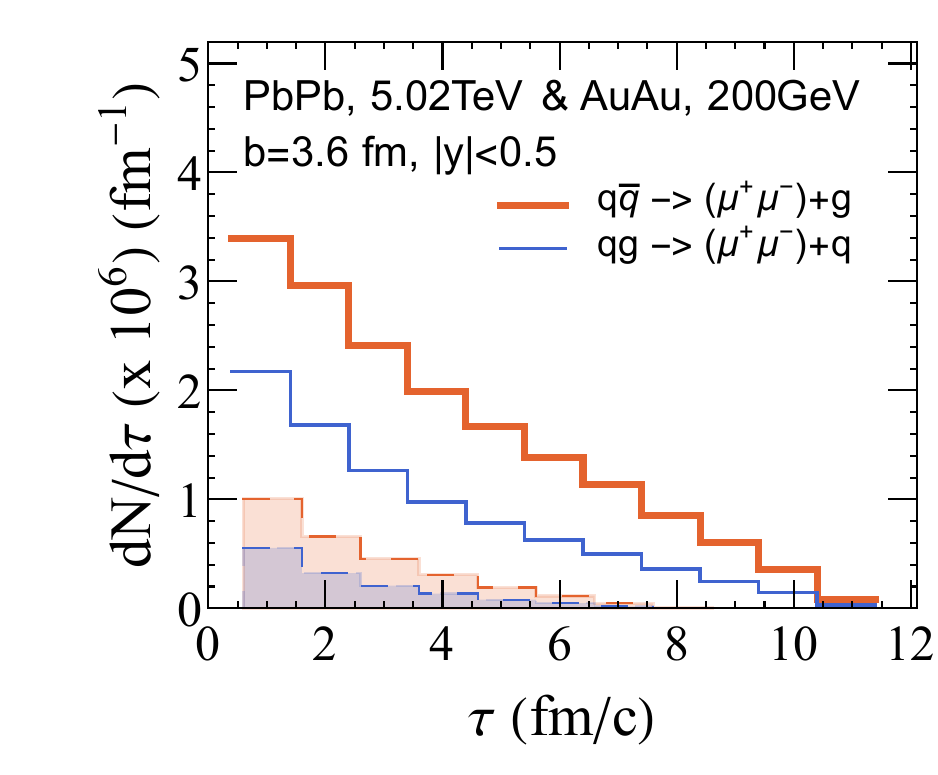}
\end{center}
\caption{The true muonium yield at each proper time $\tau$ step in the central rapidity of Au+Au collisions at RHIC energy $\sqrt{s_{\rm NN}}$ = 200 GeV (dashed) and Pb+Pb collisions at LHC energy $\sqrt{s_{\rm NN}}$ = 5.02 TeV (solid), both with impact parameter $b=\rm 3.6 ~fm$. The orange and blue curves represent the $q\bar q\to (\mu^+\mu^-)+g$ and $q g\to (\mu^+\mu^-)+q$ channels, respectively.}
\label{fig.yield-tau}
\end{figure}
%--------------------------end------------------------------------

The dynamical expansion of the quark-gluon plasma created can be effectively simulated using hydrodynamic equations in conjunction with an appropriate equation of state (EoS). In this study, we utilize the MUSIC package~\cite{Schenke:2010rr,Schenke:2010nt} to simulate the expansion of hot medium and adopt the parametrization of the lattice QCD EoS with a smooth crossover between the QGP and the hadron resonance gas around critical temperature $T_c=160 ~ \rm MeV$ ~\cite{Bernhard:2016tnd}. The initial temperature profiles of the hot medium were determined by analyzing the final multiplicity of charged hadrons. For the most central collisions and the central rapidity region, the maximum initial temperature of the medium was calculated to be $T_{\rm max}(\tau_0=0.6 ~ \rm fm/c,{\bf x}_T=0) = 320$ MeV and $T_{\rm max}(\tau_0=0.4 ~ \rm fm/c,{\bf x}_T=0) = 450$ MeV at the center of the medium in $\sqrt{s_{\rm NN}}=200$ GeV Au+Au collisions and $\sqrt{s_{\rm NN}}=5.02$ TeV Pb+Pb collisions, respectively. Here, $\tau_0$ refers to the start time of the hydrodynamics. An effective shear viscosity $\eta/s = 0.08$ and a zero bulk viscosity are chosen in this study~\cite{Bernhard:2016tnd}.

{\it Results.}
The yield and transverse momentum distribution of true muonium in Au+Au collisions at $\sqrt{s_{\rm NN}}$ = 200 GeV and Pb+Pb collisions at $\sqrt{s_{\rm NN}}$ = 5.02 TeV are shown in Fig. 4. In both cases, the $q\bar q\to (\mu^+\mu^-)+g$ channel dominates at low transverse momenta. From the yield, we define the effective production cross-section of true muonium as $\sigma_{(\mu^+ \mu^-)}=N\times \sigma_{\rm NN}^{\rm inel}\times \Delta y$, where $\sigma_{\rm NN}^{\rm inel}$ is nucleus-nucleus inelastic scattering cross-section. The obtained effective production cross sections for central heavy-ion collisions are $1.23 ~ \mu b$ at RHIC and $14.2 ~ \mu b$ at the LHC. The latter is nearly seven times larger mainly due to the higher QGP temperature at the LHC. In contrast, the counterparts in ultra-peripheral AA collisions are 19~nb and 84~nb at RHIC and the LHC, which are estimated with the cross-section of the $n^1S_0$  state production in Ref.~\cite{Azevedo:2019hqp} rescaled with the $n^3S_1$ to $n^1S_0$ event ratio in Ref.~\cite{Ginzburg:1998df}. Thus the true muonium production is significantly enhanced by approximately $\mathcal{O}(100)$ in the presence of the QGP. In fact, the obtained effective cross-section for true muonium is comparable to that of $\Upsilon$, which is roughly three orders of magnitude smaller than the cross-section of $J/\psi$. While taking into account the branching ratio to di-leptons, the yield of true muonium in heavy-ion collisions would lie between $J/\psi$ and $\Upsilon$ when detected via di-lepton reconstruction. Our calculations suggest that heavy-ion collisions (HIC) may offer the most promising opportunity for the first observation of true muonium. 

The time-dependent production of true muonium is shown in Fig.~\ref{fig.yield-tau}, which reveals that a dominant fraction is created at the early stage of the QGP. A quantitative study shows that the true muonium yield changes by a factor of two when the QGP temperature varies by 30 MeV. This behavior makes true muonium a sensitive probe for constraining the early-time temperature of the QGP. In heavy-ion collisions, direct photons serve as the primary probe for extracting the QGP temperature~\cite{PHENIX:2022rsx,Shen:2013vja}. Nonetheless, they are mostly produced at the later stages of the QGP and are significantly influenced by the collective flow, which reduces their sensitivity. Thermal dilepton methods, while useful, suffer from background contamination, particularly from open-charm hadron semi-leptonic decays, complicating the extraction of precise temperature information~\cite{Rapp:2014hha}. True muonium, predominantly produced at the early stage of the QGP, experiences less influence from flow and features a distinct resonance peak with minimal background contamination. This makes it a novel and sensitive probe of the QGP temperature at the early stage, further enhancing our ability to determine the QGP temperature with precision.

{\it Summary.} 
In this letter, we investigate the production of true muonium in relativistic heavy-ion collisions. The formation of true muonium in the QGP is described by a transport equation that includes a gain term accounting for both the $q\bar q\to (\mu^+\mu^-)+g$ and $q g\to (\mu^+\mu^-)+q$ channels. The space-time evolution of the QGP is governed by hydrodynamic equations. By solving the coupled transport and hydrodynamic equations for the relativistic heavy-ion collisions at RHIC and the LHC energies, we find that the yield of true muonium is significantly enhanced. Its sensitivity to the early QGP temperature and resilience to flow effects establish true muonium as a new powerful probe for QCD studies.

%%%%%%%%%%%%%%%%%%%%%%%%%%%%%
%\begin{acknowledgements}
%We thank...
%\end{acknowledgements}
\textbf{Acknowledgment.--}  We thank Y.Q. Ma and S. Lin for helpful discussions. ZTL is supported by the National Natural Science Foundation of China (NSFC) (Grant No.~12375075) and Shandong Province Natural Science Foundation (Grant No.~ZFJH202303). 
QY is supported by the NSFC (Grant No.~12105155) and National Key R\&D Program of China (Grant~No. 2022YFA1604903).
HZ is supported by the NSFC (Grant No.~12447105).
JXZ is supported by the Helmholtz Research Academy Hesse for FAIR (HFHF), GSI Helmholtz Center for Heavy Ion Physics.
PFZ is supported by the Guangdong Major Project of Basic and Applied Basic Research (Grant No.~2020B0301030008).

%%%%%%%%%%%%%%%%%%%%%%%%%%%%%

\end{document}